\documentclass[prd,showpacs,nofootinbib,preprintnumbers,
preprint
]{revtex4}
\usepackage{amsmath}  \usepackage{amsfonts}
\usepackage{graphicx,
xcolor
}
\def\be{\begin{equation}}
\def\ee{\end{equation}}
\def\ba{\begin{eqnarray}}
\def\ea{\end{eqnarray}}

\def\lb{\label}

\def\la{\lambda}

\def\lb{\label}
\def\W{W}
\def\X{{\hat g}}
\def\R{{\hat R}}
\def\s{{\hat s}}
\def\hnabla{\hat{\nabla}}
\def\hGamma{\hat{\Gamma}}
\def\fl{\Lambda}
\def\fL{\lambda}

\def\hBox{{\hat \Box}}

\begin{document}
\title{
Ghost-free Palatini derivative scalar-tensor theory: \\desingularization and the speed test }
\author{Dmitri Gal'tsov} \email{galtsov@phys.msu.ru}
\affiliation{Faculty of Physics,
Moscow State University, 119899, Moscow, Russia,\\Kazan Federal University, 420008 Kazan, Russia}
\author{Sophia Zhidkova} \email{zhidkova.physics137@gmail.com}
\affiliation{Faculty of Physics,
Moscow State University, 119899, Moscow, Russia}

\begin{abstract}
We disclose remarkable features of the scalar-tensor theory with the derivative coupling of the scalar field to the curvature in the Palatini formalism. Using disformal transformations, we show that this theory is free from Otrogradski ghosts.  For a special relation between two coupling constants, it is  disformally dual to the  Einstein gravity minimally coupled to the  scalar, which opens the way to constructing several exact solutions. The disformal transformation   degenerates at the boundary of the physical region, near which the desingularization properties   are revealed, illustrated by exact solutions: non-singular accelerating cosmology and a static spherically symmetric geon. We also construct the exact pp-waves of this theory propagating at the speed of light.
\end{abstract}
\maketitle


\noindent {\textbf{\textit {Introduction.}}}
   Multimessenger gravitational wave astronomy, which commenced with the detection of the binary neutron star merger GW170817 by LIGO-VIRGO collaboration \cite{TheLIGOScientific:2017qsa} and subsequent observation of its electromagnetic counterparts \cite{BPA}, showed that the velocity of gravitational waves is equal to the speed of light to within $10^{-15}$. This discovery,  anticipated in \cite{Lombriser:2015sxa}, has already had a major impact \cite{Creminelli:2017sry,Sakstein:2017xjx,Ezquiaga:2017ekz, Baker:2017hug,Langlois:2017dyl,Peirone:2017vcq} on modified theories of gravity \cite{Sotiriou:2008rp,Capozziello:2011et,Nojiri:2017ncd,Barack:2018yly,Heisenberg:2018vsk} partially refuting the most popular Horndeski (covariant Galileon) \cite{Horndeski, Deffayet,Charmousis:2011bf}  and beyond Horndeski models \cite{Zumalacarregui:2013pma,Gleyzes:2014qga,Babichev:2017guv}. At present, the restrictions imposed by the speed of gravitational waves are more stringent than  traditional cosmological bounds.   Therefore, one is urged to revise the list of existing models and look for new ones that could pass this  test.

The extended theories of gravity studied over the last decade  \cite{Barack:2018yly,Heisenberg:2018vsk}   included the conventional second-order theories with the metric connection, the first order  theories (Palatini) with an independent connection \cite{Sotiriou:2005hu,Harko:2010hw,Olmo:2011uz,Luo:2014eda,Davydov:2017kxz}, and the hybrid models \cite{Capozziello:2015lza,Olmo:2009xy}. If in Einstein's and some modified theories  both  formalisms are equivalent, this is not so in the derivatively coupled scalar-tensor theories \cite{Olmo:2009xy,Bettoni:2013diz}.  One of the major problems is to avoid the Ostrogradski ghosts. While  the Horndeski models are free from ghosts in the metric approach, they can have ghosts in the Palatini formalism. Recently this problem was  investigated using Bekenstein's disformal transformations \cite{Bekenstein:1992pj}, which turned out to be extremely useful both in the context of the metric \cite{Bettoni:2013diz,Domenech:2015hka,Sakstein:2015jca,Achour:2016rkg} and   Palatini \cite{Olmo:2009xy,Luo:2014eda} theories. These transformations, which depend on the derivatives of the scalar field, are not point-like. Nevertheless, if they are invertible, two disformally dual theories are classically equivalent \cite{Exirifard:2007da,Deruelle:2014zza,Tsujikawa:2014uza,Arroja:2015wpa,Domenech:2015tca,deRham:2016wji,Takahashi:2017zgr}.

Here we revisit the two-parameter scalar-tensor theory with derivative couplings to the Ricci tensor and  scalar  \cite{Amendola:1993uh,Capozziello:1999xt}, which  in the metric version is able to provide an inflationary mechanism without the potential  \cite{Sushkov:2009hk,Granda:2010hb}, in the case when the scalar couples to the Einstein tensor. Unfortunately,  this  model is now  in question.  We study here the derivatively coupled theory in the {\em first order} formalism, essentially using  the disformal transformation tools (this theory  was also discussed recently  for a special choice of constants \cite{Gumjudpai:2016ioy}). We show that Palatini version can be converted into a theory without higher derivatives,  proving that it is ghost-free. For some  combination of  coupling constants (other than those mentioned above) it is dual to the Einstein theory  minimally coupled to the scalar.

As a rule, the disformal transformation degenerates on a certain hypersurface in spacetime, which we interpret as the boundary of the physical region. We will show that, near this boundary, the disformal map desingularizes the metrics of the Einstein-scalar singular solutions, such as Zel'dovich stiff matter cosmology and the Fisher-Janis-Newman-Winicour (FJNW) black hole, leading to a non-singular accelerating Universe and a static spherically symmetric geon, respectively. Using the inverse disformal transformation, we also construct a  pp-wave exact solution passing the speed test.
 \smallskip

\noindent {\textbf{\textit {Metric theory.}}}
We consider the action with two independent couplings of the derivatives of the scalar field  $\phi_{\mu}\equiv\phi_{,\mu}$ to the Ricci tensor and scalar \cite{Amendola:1993uh,Capozziello:1999xt,Sushkov:2009hk,Granda:2010hb}:\vspace{-1.4mm}
\begin{equation}\lb{action}
  S\!  =\!\!\int \!\! d^{4}x\sqrt{-g}\left[R\! -\! \left(g_{\mu\nu}\! +\! \kappa_{1} g_{\mu\nu}R\! +\! \kappa_{2}R_{\mu\nu}\right)\phi^\mu\phi^\nu\right],
\end{equation}
where $\phi^\mu=\phi_\nu g^{\mu\nu}$.
 In the metric formalism, variation of  (\ref{action}) over the metric gives the equation  $G_{\mu\nu}= \Theta_{\mu\nu}$, where  $G_{\mu\nu}$ is the Einstein tensor and the right hand side contains, among others, the third  derivative terms:
\be \lb{third}
 \Theta_{\mu\nu}^{(3)} =(\kappa_2+2\kappa_1) \left( g_{\mu\nu}  \phi^{\alpha} \nabla_{\alpha}\Box\phi -\phi^\alpha\phi_{\alpha\mu\nu}\right),
\ee
where  $\nabla_\lambda$ is covariant derivative with respect to the Levi-Civita connection of $g_{\mu\nu}$,  $\Box=\nabla_\lambda\nabla^\lambda$,   $\phi_{\mu\nu}=\nabla_\mu\phi_\nu,\, \phi_{\alpha\mu\nu}=\nabla_\alpha\phi_{\mu\nu}$. Similarly,  the scalar  equation
\be
 g^{\mu\nu}\phi_{\mu\nu} +\nabla_{\mu}\left[\phi_\nu (\kappa_{1}g^{\mu\nu}R+\kappa_{2}R^{\mu\nu})\right]=0,
 \ee
in the general case contains the third derivatives of the metric. But, for  $-2\kappa_{1}=\kappa_{2}$,  the  Ricci-terms are combined into the Einstein tensor satisfying  $\nabla_{\mu}G^{\mu\nu}=0$. Then the scalar equations becomes the second order
 $ \left(g^{\mu\nu}+\kappa G^{\mu\nu}\right)\phi_{\mu\nu}=0,
$
while (\ref{third}) disappears. In this case, the theory belongs to ghost-free Horndeski class.

 \smallskip

\noindent {\textbf{\textit {Palatini theory.}}}
 The same  derivative scalar-tensor theory in the Palatini version (hereinafter abbreviated DSTP)  has  the action $S =\int d^{4}x\sqrt{-g} \,L ,$
\be  \lb{actionP}
 \! \! L\!=\!
 (\R_{\mu\nu}\!\! -\!\phi_\mu\phi_\nu) g^{\mu\nu}\!\!-\!\R_{\alpha\beta} \phi_\mu\phi_\nu (\kappa_1   g^{\alpha\beta}\! \!g^{\mu\nu} \!\! +\kappa_{2}   g^{\alpha\mu}g^{\beta\nu} )\!\!\!
\ee
where the Ricci tensor $\R_{\mu\nu}$ is a function of the independent connection $\hGamma^\lambda_{\mu\nu}$, and the Ricci scalar $\R=R_{\mu\nu} g^{\mu\nu}$ depends on the metric and on the connection.
 In absence of fermions, when the Ricci tensor contracts with symmetric tensors,   torsion can be consistently set equal to zero \cite{Afonso:2017bxr}, so the variation of $\R_{\mu\nu} $ equals
$
 \delta \R_{\mu\nu}=\hnabla_\lambda \delta\hGamma^\lambda_{\mu\nu}- \hnabla_\nu \delta\hGamma^\lambda_{\mu\lambda} ,
$
where $\hnabla_\lambda\equiv \hnabla_\lambda(\hGamma)$ denotes the covariant derivative with respect to the Palatini connection. Applying this to (\ref{actionP}), we arrive at:
\begin{align}\lb{eqW}
 &\hnabla_\lambda \left( \sqrt{-g}\W^{\mu\nu}\right) = 0,  \;\; \W^{\mu\nu}
  =\fL g^{\mu\nu} - \kappa_2\phi^\mu\phi^\nu,\\ \lb{l}
&  \fL = (1 - \kappa_1 \psi),\;\; \psi= \phi_\alpha\phi_\beta g^{\alpha\beta}. \end{align}
Variation of the action (\ref{actionP}) with respect to the metric leads to the Einstein-Palatini equation (note symmetrization in the $\kappa_2$-term, missed in \cite{Gumjudpai:2016ioy})
\be \lb{PaE}
\fL\R_{\mu\nu}-\phi_\mu\phi_\nu(1+\kappa_1 \R)  -2\kappa_2 \R_{\alpha(\mu}\phi_{\nu)}\phi^\alpha- g_{\mu\nu}  L/2 = 0.
\ee
Finally,  a variation over $\phi$ gives rise to a scalar equation
\be\lb{sp}
\partial_\mu\left[ \sqrt{-g} \left( \phi^\mu+\kappa_1 \R\phi^\mu+\kappa_2
\R_{\alpha\beta}g^{\beta\mu}\phi^\alpha\right)\right]=0,
\ee
which, in principle,  can contain  higher-derivative terms. But first  we have to build a Palatini connection.

The standard way to find the connection $\hGamma$ from the Eq.~(\ref{eqW}), is to convert it to the equation \be\lb{lc}\hnabla_\la \X_{\mu\nu}=0\ee for some second metric  $\X_{\mu\nu}$, and in this case  $\hGamma$ can be identified with the Levi-Civita connection of the latter. For this it is sufficient to ensure the following relation between the matrix  $W^{\mu\nu}$ and the inverse  metric $\X^{\mu\nu}$:
\be\lb{XW}
\sqrt{-g} \W^{\mu\nu}   =\sqrt{-\X} \X^{\mu\nu},\quad \X=\det(\X_{\mu\nu}).
\ee
 Then we get the equation in terms of the inverse new metric, equivalent to what we are looking for.

 The matrix $\W_{\mu\nu}$, an inverse of  (\ref{eqW}), reads:
 \begin{align}\lb{Wd}
&  \W_{\mu\nu}=\fL^{-1} \left(g_{\mu\nu} +\kappa_2\fl^{-1}\phi_\mu\phi_\nu\right),\\ & \fl=1 -(\kappa_1+\kappa_2)\psi.\lb{L}
 \end{align}
The determimants of the left and right sides of (\ref{XW}) are
 \be\lb{det}
 \X=g\,\fl \fL^3.
 \ee
Using this, the second metric can be represented as
 \be\lb{dsf}
\X_{\mu\nu}=\sqrt{\fl\fL}\left(g_{\mu\nu}+\kappa_2 \fl^{-1} \phi_\mu \phi_\nu  \right),
 \ee
 and  the Palatini connection, satisfying (\ref{lc}) by virtue of the Eqs. (\ref{eqW}, \ref{XW}),  will read:
 \be
 \hGamma^\lambda_{\mu\nu}= \X^{\la\tau}\left(\partial_\mu \X_{\la\nu}+ \partial_\mu\X_{\mu\la}-\partial_\la \X_{\mu\nu}\right)/2.
 \ee

 \noindent {\textbf{\textit {Disformal duality and absence of ghosts.}}}
The Eq. (\ref{dsf}) is the disformal transformation expressing the second metric $\X_{\mu\nu}$ through the physical metric $g_{\mu\nu}$, to which couples the matter. Not only the Palatini connection is nicely expressed in terms of the dual metric $\X_{\mu\nu}$, but the full DSTP theory will have a simpler form in the dual variables. Disformal dualities were extensively discussed recently  \cite{Bettoni:2013diz,Sakstein:2015jca,Domenech:2015hka,Domenech:2015tca,deRham:2016wji,Takahashi:2017zgr}. Crucial question is whether  two theories related by a non-pointlike transformation  are classically equivalent, with no extra degrees of freedom. A number of investigations  \cite{Exirifard:2007da,Deruelle:2014zza,Tsujikawa:2014uza,Domenech:2015hka,Arroja:2015wpa,Domenech:2015tca,deRham:2016wji,Takahashi:2017zgr} suggest that it is  indeed the case, provided the disformal transformation is invertible.

Expressing the Lagrangian  (\ref{actionP}) in terms of the new metric for generic $\kappa_1,\,\kappa_2$, we find  the Einstein-Hilbert term kinetically coupled to $\phi$:
 \be\lb{MES12}
S=\int \sqrt{-\X}\left[R_{\mu\nu}(\X)-\phi_\mu\phi_\nu\,{\hat \fl}^{-1}
\right]\X^{\mu\nu} d^4x,
 \ee
 where we denoted by ${\hat \fl}$ the scale factor $\fl $ with
 $\psi=\phi_\mu\phi_\nu g^{\mu\nu}$ expressed through $\hat{\psi}=\phi_\mu\phi_\nu \X^{\mu\nu}$. These two quanitities are related by an equation
 \be\lb{psipsilin}
\hat{\psi}= \psi(1-\kappa_1\psi)^{-3/2}[1-(\kappa_1+\kappa_2)\psi]^{1/2} ,
 \ee
 which is obtained combining the Eqs. (\ref{eqW}), (\ref{XW}) and (\ref{det}).
Note that we are  working in the Palatini formalism, so the Ricci tensor in (\ref{MES12}) should  be considered as a functional of the connection. But for such an action both the metric and the Palatini approach give  the same equations, so, with some abuse in the notation,  we have denoted the Ricci tensor is as metric.  Therefore, we reduced the dynamics of the scalar field to the problem of Einstein's theory. After solving it, we can restore the DSTP metric using an algebraic procedure. 
 
 The relation (\ref{psipsilin})  becomes singular if one of the scale factors (\ref{l}),  (\ref{L})  reaches zero,
 when the determinant ratio (\ref{det}) degenerates. We thus demand  positivity of $\fL , \fl $ in the physical region of spacetime. Also, the transformation must preserve the metric signature. The necessary condition for this  is  the common sign of the norms $\hat{\psi}$ and $\psi$. This is guaranteed by the Eq.~(\ref{psipsilin}) if  $\fL>0, \fl>0$. 
 
 A more subtle question is whether the mapping is one-to-one. Generically, it is not. The derivative  $d{\hat \psi}/d\psi=0 $   for $\psi=\psi_{\rm cr}=2/(2\kappa_1+3\kappa_2)$. The corresponding dual norm is
$
 {\hat \psi}(\psi_{\rm cr})=2/(3\sqrt{3}\kappa_2).
$
To the right and to the left of this point, the function ${\hat \psi}(\psi)$ is monotonous (the left panel in Fig. 1), so the map is one-to-one.
But the inverse derivative $ d\psi/d{\hat \psi}$ diverges at the critical point. As a consequence, the Einstein-frame equations become singular there. But, it is expected that  viable solutions  of the Einstein-frame equations will be determined in a physical region  where the critical point is not met. At least, this is true for a sub-family of the theories described below.  With these precautions, we can  assert that our disformal map is reversible, in contrast to what happens in mimetic gravity~ \cite{Chamseddine:2013kea,Barvinsky:2013mea,Sebastiani:2016ras}. 
\begin{figure}[ht]\center
\includegraphics[width=140mm]{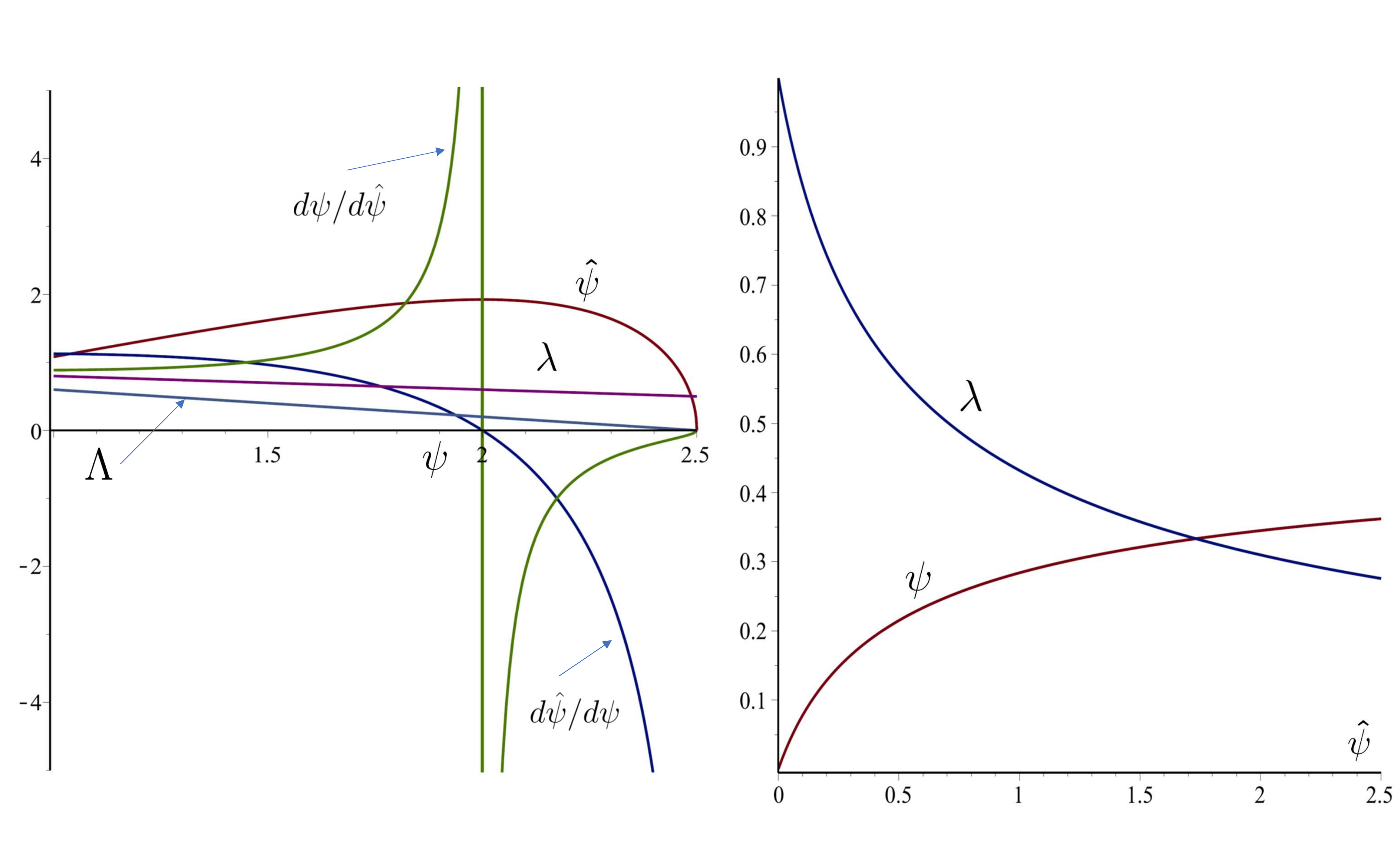}
\caption{Left panel: dependence ${\hat \psi}(\psi)$ (\ref{psipsilin}), derivatives  $ d{\hat \psi}/d\psi,\; d\psi/d{\hat \psi}$ and scale factors
 $\lambda({\psi})$, $\Lambda({\psi})$ for generic theory ($\kappa_1=\kappa_2=0.2$. Right panel: solution $\psi({\hat \psi})$ of Eq. (\ref{psipsilin}) and scale factor $\lambda$ for exceptional theory ($\kappa_1=-\kappa_2=1$).}
\end{figure}
Since the theory (\ref{MES12}) does not  contain higher derivatives, we conclude that the DSTP theory with two generic coupling constants  does not have the Ostrogradski ghosts.

 \smallskip

 \noindent {\textbf{\textit {Exceptional  theory.}}}
 The case $\kappa_2=-\kappa_1 $ is exceptional. Then $\fl\equiv 1$, and the theory (\ref{MES12})   reduces  to 
  Einstein theory,  minimally coupled to a massless scalar:
 \be\lb{MES}
S=\int \sqrt{-\X}\left[R_{\mu\nu}(\X)-\phi_\mu\phi_\nu\right]\X^{\mu\nu} d^4x.
 \ee
 In this dual theory the Einstein equation reads
 \be\lb{EX}
 R_{\mu\nu}=\phi_\mu\phi_\nu,
 \ee
 and the scalar obeys the covariant d'Alembert equation
\be\lb{sb}
{\hat\Box}\phi=0,
\ee

One can verify that Eqs. (\ref{PaE}) and (\ref{sp}) are satisfied by virtue of Eqs. (\ref{EX}) and (\ref{sb}). First, we obtain that Eq. (\ref{EX}) implies
$ L = 0, \, \R = \psi $, hence Eq. (\ref{PaE}) holds. Using then Eq. (\ref{EX}) in Eq. (\ref{sp}), we reduce the latter to (\ref{sb}).

The DSTP metric $g_{\mu\nu}$ can be found from the Einstein metric $\X_{\mu\nu}$ and the scalar $\phi$ via Eq. (\ref{dsf}), giving
\begin{align}\lb{gmet}
& g_{\mu\nu} = \X_{\mu\nu}\fL^{-1/2}+\kappa_1\phi_\mu\phi_\nu \\ \lb{lamb}
&\!\!\!\!\!\fL^{1/2}= \frac{\sqrt{3}}{2z}\begin{cases}
  2\cos\left(\frac13\arccos(2z^2-1)\right)-1, &\!\!z<1,\\
A^{1/3} +A^{-1/3}-1, & \!\!z>1,
 \end{cases}\end{align}
 where $A=2z\sqrt{z^2-1}+2z^2-1,\; z=3\sqrt{3}\kappa_1{\hat \psi}/2$. The function (\ref{lamb}) is smooth at $z=1$ (the right panel in Fig. 1). Here  $\kappa_1, \psi$ and ${\hat\psi}$ are chosen positive (spacelike $\phi_\mu$). The critical point of the map (\ref{psipsilin}) in this case lies in the region of negative $\psi, {\hat\psi}$, which is not shown.  The cases of timelike and null $\phi_\mu$ are considered in the concrete examples below. In all cases the sign of the norm of $\phi_\mu$ is preserved within the physical region.   
\smallskip

\noindent {\textbf{\textit {Non-singular cosmology.}}}
Assume that $\kappa_2=-\kappa_1\equiv\kappa>0$. Using disformal duality, we can construct an exact cosmological solutions of our theory starting with the spatially flat FRW cosmology in the Einstein's frame theory (\ref{MES})
\be\lb{frw}
d\s^2=\X_{\mu\nu} dx^\mu dx^\nu=-dt^2+{\hat a}^2\delta_{ij}dx^i dx^j.
\ee
The relevant Einstein and scalar field equations
\be
R_{tt}=- \frac{3\ddot{{\hat a}}}{{\hat a}}={\dot\phi}^2,\quad {\ddot \phi}+3\frac{\dot{{\hat a}}}{{\hat a}}\dot{\phi}=0
\ee
give a solution corresponding to the ``stiff-matter'' cosmology proposed by Zel'dovich in 1972 \cite{Zeldovich,Chavanis:2014lra}:
\be\lb{Ze}
{\hat a}=a_0 t^{1/3},\quad \phi=\sqrt{ 2 }\ln t/\sqrt{3}.
\ee
Obviously, this metric is singular at $t=0$ and describes a decelerating expansion.

Now we derive the corresponding solution to DSTP theory. Since the metric is diagonal and the scalar field depends only on  $t$, from (\ref{gmet}), (\ref{frw}) and (\ref{Ze}) we obtain  a cubic algebraic equation for $g_{tt}$:
\be
\left(|g_{tt}|-2x/(3\sqrt{3})\right)^3=|g_{tt}|,\quad x=\kappa \sqrt{3}/t^2.
\ee
Its real solution is   smooth, although in terms of real functions it looks piecewise:
\be\lb{gtt}
|g_{tt}| = \frac{2x}{3\sqrt{3}}+\frac1{\sqrt{3}}
 \begin{cases}
  2\cos\left(\frac13\arccos(x)\right), &x<1,\\
B^{1/3} +B^{-1/3}, & x>1,
 \end{cases}
\ee
where $B=\left(x+\sqrt{x^2-1}\right)^{1/3}$.
For large $x$ (small $t$) one has:
\be\lb{sing}
|g_{tt}| =  {2x}/{3\sqrt{3}}+{(2x)^{\frac13}}/{\sqrt{3}}+\left(4/{x}\right)^{\frac13}/{(2\sqrt{3})}+...\,,
\ee
for small $x$ (large $t$),
\be\lb{as}
|g_{tt}| =1+ {x}/{\sqrt{3}}- {x^2}/{18}+...\;.
\ee
For $g_{ij}$  one obtains from (\ref{gmet}):
\be
g_{ij}=\delta_{ij} a^2,\quad  a^2={\hat a}^2  |g_{tt}|^{1/3}.
\ee

Since $|g_{tt}|=1$ only at large $t$, we need to go to the synchronous time $t\to\tau(t)$, so that $g_{\tau\tau}\equiv 1$, solving the equation
\be
dt/d\tau=|g_{tt}|^{-1/2}.
\ee
For small $t$, keeping the leading term in (\ref{sing}), one finds:
\be
dt/d\tau=H_0 \,t \,\Longrightarrow \, t={\rm e}^{H_0 \tau},\,H_0=\sqrt{ 3/{(2\kappa)}}.
\ee

For spatial components at small $t$, with account for (\ref{Ze}) and  (\ref{sing}), we obtain: $a^2\to a_0^2 H_0^{-2/3} $. Rescaling  spatial coordinates, we get the Minkowski metric as $t\to 0$.  Calculating   $\ddot{a}$, we see that the universe starts from Minkowski stage with an acceleration which, however, ends   with a small gain of $a$ (Fig. 2).
\begin{figure}[ht]\center
\includegraphics[width=100mm]{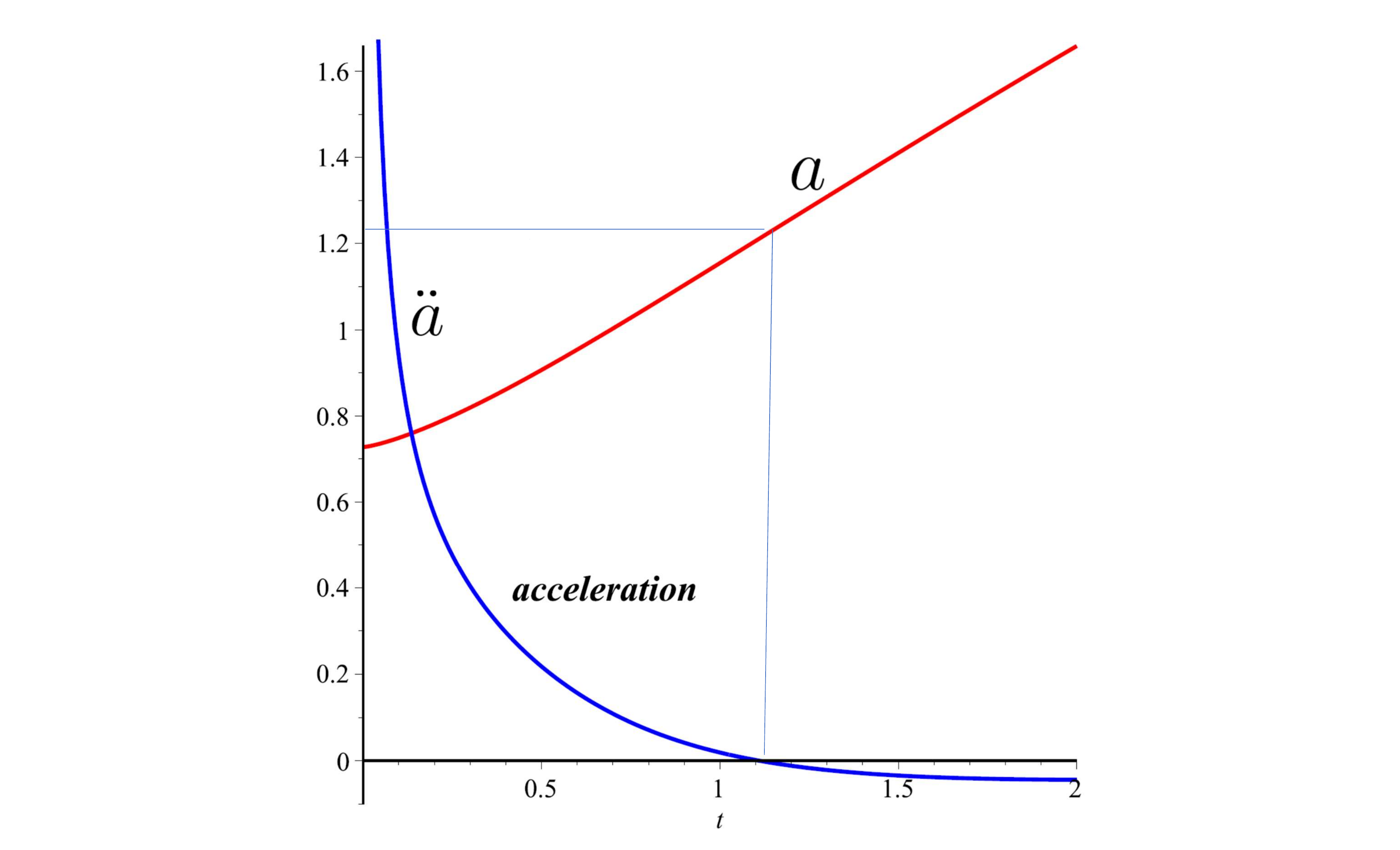}
\caption{Non-singular DSTP cosmology. The expansion begins with a finite scale factor with positive acceleration.}
\end{figure}
Then expansion decelerates and at large $t$ the  law $a\sim t^{1/3}$ is restored, since  $\la\to 1$ and the comoving time coincides with the Einstein frame comoving time.

Although the DSTP metric  is nonsingular as $\tau\to -\infty$,  the scalar field  diverges. In  Palatini theories  it is assumed that matter couples to the metric $g_{\mu\nu}$, so  geodesics, defined as curves of minimal length, do not stop as $\tau\to -\infty$ and the spacetime is geodesically complete. But, the auto-parallel curves, defined with the Levi-Civita connection of the singular metric (\ref{frw}),  (\ref{Ze}),  meet the singularity. Still, if matter couples only to the metric, the scalar field is not seen. It appears as Deus Ex Machina, realizing the cherished dream of the General Relativity --- the removal of singularities from spacetime.

 \smallskip

 \noindent {\textbf{\textit {Geon.}}}
In the static case, interesting solutions arise for  $\kappa_1=-\kappa_2>0$, so here we denote $\kappa =\kappa_1 $. Minimal scalar gravity has a well-known FJNW solution \cite{Fisher,JNW}
\begin{align}
& \X_{tt}=-\X^{-1}_{rr}=-\left(1-{b}/{r}\right)^{\gamma},   \; \X_{\theta\theta}=r^{2}\left(1-{b}/{r}\right)^{1-\gamma},\nonumber\\
& \phi= {q}(b)^{-1}\ln\left(1-{b}/{r}\right),
\end{align}
\noindent where $q$ is the scalar charge and $0<\gamma<1, \;\gamma=\left(1- {4q^{2}}/{b^{2}}\right)^{1/2}$. It is asymptotically flat and has a singular horizon at $r=b$.
We want to find a DSTP counterpart of this solution. The disformal transformation (\ref{gmet}) generates now the following cubic equation for $g_{rr}$:
\begin{align}
&[g_{rr}-2x/(3\sqrt{3})]^3=w^2 g_{rr},\;\;\; w=\X_{rr}=\left( 1-b/r\right)^{-\gamma},\nonumber\\ & x=3\sqrt{3} \kappa q^2/[2 r^2 (r-b)^2]
\end{align}
A smooth real solution of this equation is:
\be\lb{gtt}
g_{rr} \! \!= \! \!\frac{2x}{3\sqrt{3}}\! \!+\frac1{\sqrt{3}}\! \!
 \begin{cases}
  2w\cos[\frac13\arccos(x/w)] ,&    x<w,\\
  w^{2/3}B+w^{4/3}B^{-1} , &  x>w,
 \end{cases}
\ee
where $B=\left(x+\sqrt{x^2-w^2}\right)^{1/3}$. The remaining metric components then are given by:
\be
 g_{tt}\!=\! {\X_{tt}}/{\fL^{1/2}}\!,\,g_{\theta\theta}\!=\! {\X_{\theta\theta}}/{\fL^{1/2}},\,\fL=\left(g_{rr}/w \right)^{-2/3}.
 \ee

At infinity $r\to\infty$, the variables $x\to 0,\;w\to 1$, while $g_{rr}\sim 1+x/\sqrt{3}$, so  $\lambda=1+O(r^{-4})$ and the solution remains asymptotically flat:
\be
g_{tt}\sim -1+\gamma b/r,\quad g_{rr}\sim 1-\gamma b/r,\quad g_{\theta\theta}\sim r^2.
\ee

Near the singularity $r\to b$,  one has $w\sim \xi^{-\gamma}, x\sim \xi^{-2}$, where $\xi= r/b-1\to 0$ leading to:
\begin{align}
&\lambda\sim\mu^{-2}\xi^{2(2-\gamma)/3},\;\; g_{tt}\sim\mu \xi^{2(2\gamma-1)/3},  \;\;g_{rr}\sim\mu^3\xi^{-2},  \nonumber\\ & g_{\theta\theta}\sim\mu b^2 \xi^{(1-2\gamma)/3},\;\;\mu=(\kappa q^2/b^4)^{1/3}
\end{align}
  For $\gamma=1/2$, the interval in the vicinity of $r=b$  reads:
\be
\mu^{-1} ds^2=-dt^2+(\mu dr/\xi)^2 +b^2(d\theta^2+\sin^2\theta d\varphi^2).
\ee
Passing to the new radial coordinate $\rho=b\ln \xi$ extending the domain $ r\in (b, \infty)$ to a complete real line, one can find that the manifold  is isomorphfic to the product $M_{1,1}\times S^2$ of the  two-dimensional Minkowski space and a sphere of  radius $b$. This manifold is geodesically complete. Therefore, our solution has a metric of  a regular geon. Its striking feature, however, that it is supported by a singular scalar.
 \smallskip

  \noindent {\textbf{\textit {PP-waves.}}}
Wave spacetimes (and more general classes of Kundt metrics) in Einstein theory coupled to the minimal scalar were constructed recently \cite{Tahamtan:2015iqa}. Here we consider the simplest pp-wave 
\be
d\s^2= F(u,x,y)\; du^2-2 du dv + dx^2+dy^2,
\ee
whose tensor  Ricci is
\be
R_{\mu\nu}=-\delta^u_\mu \delta^u_\nu \,\Delta \,F/2,\quad \Delta=\partial_x^2+\partial_y^2.
\ee
 Assuming that the scalar field depends only on $u$, it is easy to find that the d'Alembert equation of the dual theory is satisfied: $\hBox \phi(u)=0$, and the Einstein equation  $R_{\mu\nu}=\phi_\mu\phi_\nu$ reduces to an equation for $F$:
 \be
 \Delta \,F=-2\phi^{\prime 2}.
 \ee

 Now construct the corresponding DSTP solution. In this case, the disformal transformation is light-like \cite{Lobo:2017bfh}. Assuming that  the   only  nonzero ones are $g_{uu},\, g_{uv},\,g_{ij}$, we find  non-zero inverse metric components  $g^{vv},\, g^{uv},\,g^{ij}$, therefore $\psi= \phi_\mu\phi_\nu g^{\mu\nu}=0$, and  the scale factor $\lambda=1$. Then from the Eq. (\ref{gmet}) we easily find the pp-wave solution of the DSTP theory:
 \be
d\s^2= \left(F(u,x,y)-\kappa \phi^{\prime 2}\right) du^2-2 du dv + dx^2+dy^2.
\ee
It is clear that it propagates at the speed of light.
 \smallskip

\noindent {\textbf{\textit {Conclusions.}}}
We have proved that the  DSTP theory  is related to Einstein's gravity, kinetically coupled to a scalar without higher derivatives by means of an invertible mapping, and hence it is free of Osrogradski ghosts. For opposite couplings $ \kappa_1 = - \kappa_2 $, the dual theory is just Einstein's theory, minimally coupled to the scalar. The inverse duality transformation is resolvable analytically, opening the way for finding exact solutions to  DCTP. We built cosmological, static spherically symmetric and pp-wave DSTP solutions with intriguing properties: the cosmological solution has a completely nonsingular metric, the static solution is geon-like, and the pp-wave propagates exactly at the speed of light. The scalar field diverges, taking responsibility for the singularities, but these are not visible if the matter couples only to the metric. This gives a new look at the problem of singularities in theories of gravitation.

 \noindent {\textbf{\textit {Acknowledgments.}}}
The authors are grateful to G\'erard Cl\'ement  for careful reading of the manuscript and  valuable comments. We also  thank Evgeny Babichev, Andrei Barvinsky, Salvatore Cappozziello, Jose Beltr\'an Jim\'enez, Francisco Lobo, Sergei Sushkov and Michael Volkov for  useful discussions. The work was supported by the Russian Foundation for Basic Research on the project 17-02-01299a, and by the Government program of competitive growth of the Kazan Federal University. SZh also acknowledges the support of the   Foundation for Theoretical Physics "Basis".

\end{document}